# Pulsed Frequency Shifted Feedback Laser for Accurate Long Distance Measurements: Beat Order Determination.

Jean-Paul Pique [a,*]

[a]*Univ. Grenoble 1 / CNRS, LIPhy UMR 5588, Grenoble, F-38041, France*

[*]*Corresponding author: pique@liphy.ujf-grenoble.fr, tel: +33 4 76 51 47 45, Laboratoire Interdisciplinaire de Physique, 140 rue de la physique BP87, 38402 Saint Martin d'Hères, France.*

Long-distance measurements (10 m - 1000 m) with an accuracy of $10^{-7}$ is a challenge for many applications. We show that it is achievable with Frequency Shifted Feedback (FSF) laser interferometry technique, provided that the determination of the radio frequency beat order be made without ambiguity and on a time scale compatible with atmospheric applications. Using the pulsed-FSF laser that we developed for laser guide star application, we propose and test, up to 240 m, a simple method for measuring the beat order in real time. The moving-comb and Yatsenko models are also discussed. The first of these models fails to interpret our long-distance interferometry results. We show that the accuracy of long-distance measurements depends primarily on the stabilization of the acoustic frequency of the modulator.

## 1. Introduction

Measurement of long-range distances (10 m - 1000 m) with an accuracy of $10^{-7}$ is a major challenge for the automotive, aerospace, nuclear, geology, atmospheric applications ... Commercial systems are already available that have an accuracy of the order of few ppm. In general, three categories of methods exist: triangulation, time of flight and interferometry [1]. The requisite accuracy can be provided by laser interferometry methods. Interferometry with a single laser is very accurate, but applicable only for very short distances, because the fringe size is the laser wavelength. For longer distances, the interference order to be determined becomes extremely high. This difficulty has been overcome by the "Synthetic Wavelength Interferometry" method (SWI) [2], which has demonstrated beautiful results [3, 4]. The method consists in superimposing two beams of frequency $\nu_1$ and $\nu_2$ in an interferometer. The synthetic wavelength $\Lambda = c / (\nu_2 - \nu_1)$ then lies in the millimeter range. For an optical path difference $z$, the interference order $m = \left\lfloor \dfrac{z}{\Lambda} \right\rfloor$ is determined by scanning the frequency of one laser. Distances of several tens of meters have been measured in this way with an accuracy of ppm. This method using two lasers, however, remains more difficult to implement for various technical reasons. With a single laser, the easiest technique, called "Modulated Continuous-Wave Frequency" (FMCW) [1] or "Optical Frequency Domain Reflectometry" (OFDR) [5] consists in quickly scanning the wavelength of a high spectral purity laser. The beat frequency detected at the output of the interferometer is proportional to the optical path difference, and generally lies in the kilohertz range. Unfortunately, this is in an area where the intrinsic noise of the laser is still high. Furthermore, the nonlinearity of the scan remains a strong obstacle. For all the interferometry techniques mentioned above, the maximum measurable distance is limited by the coherence of the laser line. In an experiment of great ingenuity, Nakamura et al. [6, 7] have shown that an FSF laser can measure distances up to more than 18.5 km with an accuracy of a few ppm. This new technique is similar to the OFDR method but with a beat in the radio frequency range. Although it is very promising, it nevertheless suffers from the difficulty and the slowness of determining the beat order.

In this paper we show that the pulsed-FSF laser that we developed for laser guide star [8] solves many of the problems outlined above. In the second section we describe the experimental set-up used. In the third section we describe a new technique to determine unambiguously, and in a few ms, the RF-beat order. We show experimentally that the well known moving-comb model fails to interpret the results of long distances measurements. The model of Yatsenko et al. [9] is more relevant. Finally in the fourth section, we discuss the conditions required to achieve an accuracy of $10^{-7}$.

## 2. Pulsed-FSF laser long distance measurement set-up

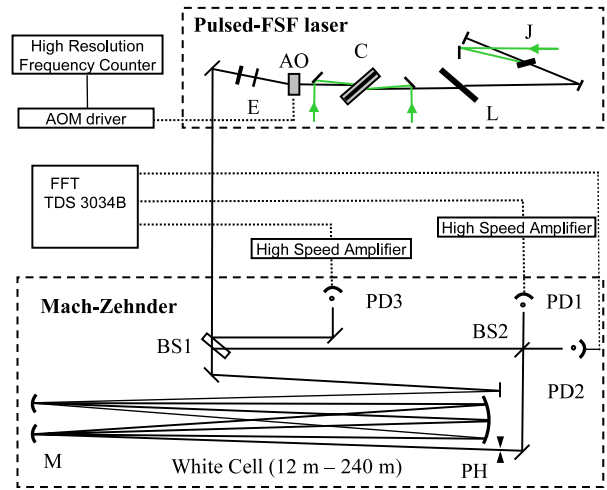

Fig. 1

The set-up is depicted in Figure 1. The pulsed-FSF laser is detailed in reference [8]. Briefly, the laser cavity consists of two dye amplifier media. The dye-jet (J) is pumped by a cw frequency doubled VANADATE laser (VERDI/Coherent) and the dye-cell (C) is pumped by a pulsed frequency doubled YAG laser operating at a repetition rate of 2 kHz with pulse duration of 35 ns (Navigator II/Spectra Physics). Both are quasi-longitudinally pumped. The acousto-optic shifter (AO) operates at an acoustic frequency $\nu_{AO}$ of 35 MHz. The cavity is closed on the first diffraction order of the AO. The only selective elements of the cavity are: a three-quartz-plate Lyot-filter (L) and an etalon (E) (with a free spectral range of 400 GHz and a 40% reflection coefficient on each side). Then, our FSF laser can have two line widths: 30 GHz and 4.5 GHz respectively. The AO also acts as a broadband filter through the grating that is generated. This arrangement yields a single line and a cavity free spectral range $\nu_{RT}$ of 253 MHz.

The use of a Michelson interferometer creates additional RF-beats and intra-cavity disturbances. A Mach-Zehnder configuration avoids feedback of the beam to the output mirror of the FSF laser. A first beam splitter (BS1) reflects two 4% beams. The intensity and power spectrum of the first beam are analyzed using a fast photodiode (PD3). The second beam is injected into the reference arm of the Mach-Zehnder. The optical path of the transmitted beam is elongated in a White cell of focal length 3 m. The output beam of the White cell is superimposed on the reference beam through a 50% beam splitter (BS2). One advantage of the White cell is the mode matching of the two beams. Use of a pinhole (PH) allows perfect superposition whatever the path difference, which is adjusted by horizontal movement of mirror M. The length of the second arm of the Mach-Zehnder can be varied easily from 12 m to 240 m in steps of approximately 12 m. Three fast photodiodes are used (rise time < 1 ns). The photodiode PD1 records the interferometric signal $I_{int}$. As will be seen later, the photodiode PD2 is used to measure the RF-beat order. The signals from PD1 and PD3 are filtered and amplified (36 dB, frequency bandwidth: 50 kHz to 1.5 GHz). The three signals are analyzed using an oscilloscope with FFT capability (Tektronix TDS 3034B) having a precision time base of $2\times10^{-5}$. The frequency of the AO shifter is measured in real time by a frequency counter with an accuracy of about $2\times10^{-7}$ (Racal Dana 1992 including a crystal oven).

## 3. New method for fast beat order determination

The emission characteristics of the FSF laser differ significantly from those of conventional lasers. A FSF laser intra-cavity optical wave $\nu_0$ is shifted to $\nu_0+2\nu_{AO}$ after a round trip through the AO shifter, where $\nu_{AO}$ denotes the frequency of the corresponding acoustic wave. This continuous change in frequency disrupts the constructive interference of the fixed modes in a conventional laser. The interest of such lasers is their very high chirp rate, which, in this work is about 20 MHz /ns.

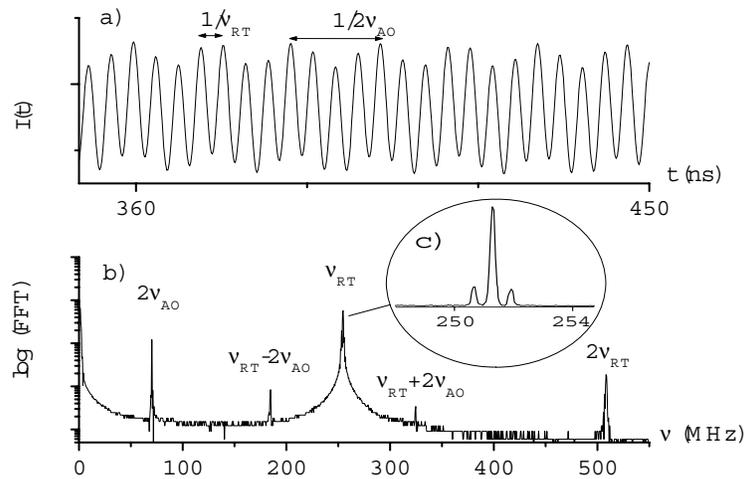

Fig.2

Figure 2a and 2b show respectively an expansion of the laser intensity $I(t)$ (PD3 signal) and the log of its FFT (PD3 signal). The recurrences $n\times \nu_{RT}$ ($n = 1,2 ...$) are clearly observed. Recurrences for $n> 1$ are strongly attenuated by the 300 MHz bandwidth of our oscilloscope.

The narrow beat frequency $2\nu_{AO}$ was attributed [10] to a residual amplitude modulation due to imperfections in the AO frequency shifter. The very weak satellites $\nu_{RT} \pm 2\nu_{AO}$ we observe (Fig. 2b) prove an amplitude modulation at a frequency of $2\nu_{AO}$ (see below). The existence of this modulation on the beam is an obstacle for experimental demonstration of a hypothetical moving-comb (see below).

Other sidebands (Fig. 2c), close to $\nu_{RT}$ recurrences correspond to an already observed amplitude modulation frequency $\nu_{am}$ of about 1.25 MHz [8]. Here we interpret these as interruptions of the intra-cavity field due to air-bubbles that flow through the dye cell and pass in the spatial mode of the intra-cavity beam. These interruptions, which occur on average every few tens of microseconds, induce relaxation oscillations just after the build-up. Based on the underlying non-linear dynamics, the frequency modulation can be satisfactorily described by the photon density model of reference [8]. These sidebands are intrinsic to the amplifying medium.

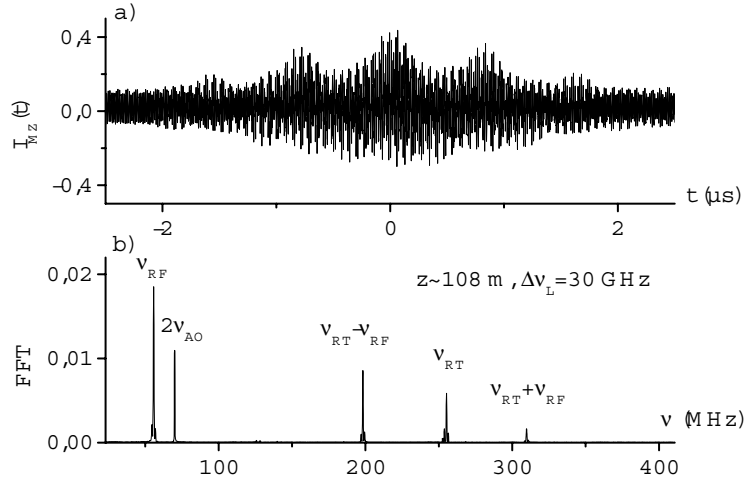

Fig.3

Figure 3 depicts the FFT of the Mach-Zehnder output signal (photodiode PD1) for a path difference $z \sim 108\ m$. Radio frequency doublets $\pm \nu_{RF}$ are observed on both sides of the $\nu_{RT}$ recurrences. The corresponding frequency difference allows accurate distance measurements (see below). Recurrences and doublets are themselves flanked on either side by the $\pm\nu_{am}$ amplitude modulation satellites.

Observation of recurrences and doublets gave rise to the moving-comb model [11], which has hitherto satisfactorily described most interferometric results in the literature. It consists of considering that the frequency comb of an equivalent multi-mode laser (without AO) shifts at the rate $2\nu_{AO}/\tau_{RT}$, where $\tau_{RT} = 1/\nu_{RT}$. To our knowledge, only two papers report the observation of moving-comb in very different experimental conditions. By injecting the FSF laser beam in a Fabry-Perot interferometer of sufficient resolution and fixed length, the passage of "modes" would produce signal peaks of frequency $2\nu_{AO}$ on a fast photodiode. This experiment is very delicate because the detection bandwidth must be between $2\nu_{A0}$ and $\nu_{RT}$. The first demonstration was made in reference [12] using an FSF dye laser having the following characteristics: $2\nu_{AO} = 0.2$ MHz and $\nu_{RT} = 188$ MHz. As expected in the moving-comb model, the time response of the signal at the output of a high resolution Fabry-Perot interferometer displays temporal peaks spaced in time by $1/2\nu_{AO}$. A second experimental demonstration was performed in references [13, 14]. An FSF Nd: YVO4 diode-pumped laser was used with the following characteristics: $2\nu_{AO} = 157$ MHz and $\nu_{RT} = 1492$ MHz. These observations seemed to validate the moving-comb model (at least when $2\nu_{AO}$ is smaller than, and even close to, $\nu_{RT}$). However, we will see below that it fails to interpret our experimental results of long-distance interferometry.

Without using the spectral aspect of the moving-comb model, Yatsenko et al. [9] introduced a more realistic stochastic model by including frequency dependent gain and losses of an FSF laser seeded by spontaneous emission. The authors calculated that the RF spectrum of the output signal of an interferometer injected by an FSF laser exhibits doublets whose spacing is directly proportional to the path difference between the two interferometer arms.

In both models the distance measured by interferometry is determined modulo of a characteristic distance $z_0$. As discussed below, $z_0$ is in metric range that, compared to SWI techniques, makes things easier for measurements over long distances. However, it is also necessary to determine the beat order.

### *3.1 Beat order in the moving-comb model*

Because $2\nu_{AO} < \nu_{RT}$ (75 MHz and 253 MHz, respectively) with a ratio close to that of the experimental work of reference [14] where the moving-comb seems to have been observed experimentally, it is tempting to use this picture. Kasahara et al. [11] proposed a model in which the field is expressed as:

$$E(t) = \sum_q E_q(t) \exp(2\pi j \Phi_q(t))$$

$$E_q(t) = E_0(t) \exp\left[-\left\{\frac{2\nu_{AO}}{\tau_{RT}}(\frac{t - q/2\nu_{AO}}{\Delta\nu_L/2})\right\}^2\right] \quad (1)$$

$$\Phi_q(t) = (t - q/2\nu_{AO})\nu_0 + (t - q/2\nu_{AO})^2 \frac{\nu_{AO}}{\tau_{RT}} + \Phi_q$$

$$E_0(t) = A(1 + B\sin(2\pi\nu_{am}t))$$

The signals at the photodiodes PD1 and PD3 are:

$$I(t) = |a_1 E(t)|^2$$
$$I_{int}(t) = |aE(t) + bE(t - \tau)|^2 \quad (2)$$

where $q$ is the index of the modes. $\Delta\nu_L$ is the spectral width of the laser line, or more precisely the width of the spectral envelope. The time delay between the two arms of the Mach-Zehnder is $\tau = z/c$, where $z$ is the optical path-difference of the two arms of the Mach-Zehnder, and $a$ and $b$ are transmission coefficients on each arm. $E_0(t)$ was added to take into account the amplitude modulation discussed above.

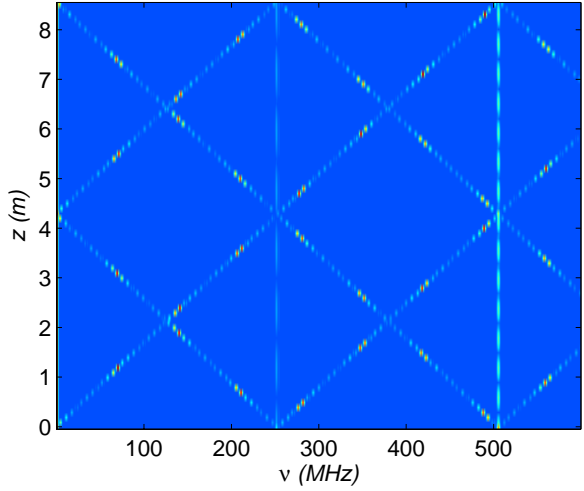

Fig. 4

Figure 4 depicts the FFT dependence of $I_{int}(t)$ on frequency $\nu$ and distance $z$. We see clearly the fixed recurrences $\nu_{RT}$ and beats $\nu_{RF}$, which are written:

$$\nu_{RF}^m = \nu_{RT}(\frac{z}{z_0} \pm m)$$
$$z_0 = \frac{c}{2\nu_{AO}}$$
(3)

where $m$ is the beat order and $z_0$ is the equivalent of the fringes in the SWI technique. With the parameters of the previous section, $z_0$ is about 4.29 m. For a distance of 100 m, the order $m$ is only 23. The power of the method is obvious: we can measure distances merely by detecting the radio frequency in the range $0$-$\nu_{RT}$ (in practice the interval $0$-$\nu_{RT}/2$ is sufficient), provided that the determination of the beat order $m$ is unambiguous. In our case the highest frequency to be measured is 253 MHz. The 300 MHz bandwidth of a relatively cheap oscilloscope is sufficient. Note that this bandwidth avoids the spurious peaks of the well known aliasing phenomenon.

## *3.2 Beat order in the Yatsenko model*

In this realistic model the electric field of the FSF laser increases from broadband spontaneous emission injected into an active laser cavity containing a gain medium, a spectral filter and an optical frequency shifter. From the spontaneous emission, the succession of round trips in the cavity generates spectra of evenly spaced $2\nu_{AO}$ frequencies. The underlying idea is similar to that used to calculate the photon density of an FSF laser from a rate equation model [15, 16]. Yatsenko calculates the second and the fourth order correlation function of the electric field $G_2$ and $G_4$ to interpret observations on the laser intensity (PD3) and the output intensity of a two beam interferometer (PD1). The authors show that $G_4$ is the sum of two terms: $G_3^{(4)}$ and $G_{int}^{(4)}$, which, in the power spectrum, respectively generate $\nu_{RT}$ recurrences and $\nu_{RF}$ doublets with a separation that depends on the path difference of the interferometer. In our notation, these terms are:

$$G_3^{(4)}(t) = G \sum_{M=-\infty}^{\infty} \frac{1}{16}(2F_M(t) + F_M(t-z/c) + F_M(t+z/c)) \quad (4)$$

$$G_{int}^{(4)}(t) = G \sum_{M=-\infty}^{\infty} F_M(t) \cos(M \frac{4\pi \nu_{AO} z}{c}) \quad (5)$$

with

$$F_M(t) = \frac{\Delta \nu_L^2}{2\pi} \left[ \exp(-\frac{(t-M\tau_{RT})^2 \Delta \nu_L^2}{4}) \right]^2 \exp(-\frac{M^2}{4M_0^2}) \quad (6)$$

$M$ is the round-trip number in the cavity. $M_0$, the number of $2\nu_{AO}$ spectral elements in the laser line, is $M_0 \sim \Delta\nu_L/2\nu_{AO}$. Note that $M_0\tau_{RT}$ is the effective lifetime of the photon in the active cavity. The cosine of equation (5) is the key term that generates doublets.

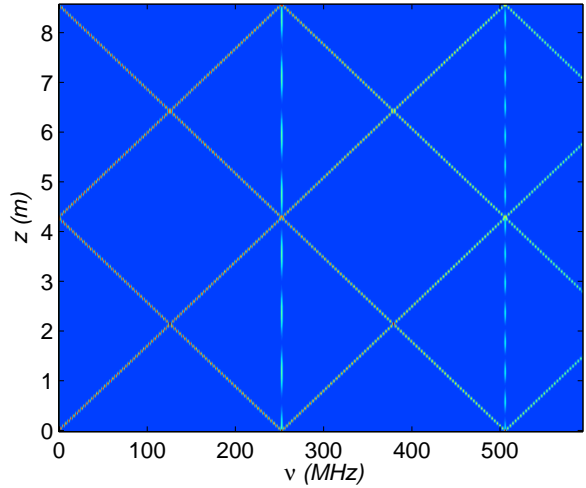

Fig. 5

Figure 5 shows the dependence of the FFT of $G_4$ on the RF frequency $\nu$ and the path difference $z$. It is quite similar to Figure 4 of the moving-comb model. The RF frequency follows the same law: it is proportional to modulo $z$ of the same characteristic length $z_0$. The beat order must therefore be determined independently. This model shows that long-range coherence is not contained in the optical field but in the properties of intra-cavity frequency shifter. A recent experiment [17], using two dye FSF lasers with identical cavities, shows that it is indeed the phase of the AOM driver that maintains coherence.

However, although the model of Yatsenko et al. is applied to an active cavity, it does not show the over intensities corresponding to resonances at frequencies $\nu_{RF} = k \times 2\nu_{AO}$ which are observed in the moving comb model (see Figure 4) and which are also observed both in the literature and in our experiments. This point is not essential insofar as we are interested only in distance measurements, but it becomes important when reflectometry measurements are considered.

### *3.3 New method for fast beat order determination*

In the framework of the moving-comb model, references [11] and [18] give a method of determining $m$. Without going into detail, it consists in measuring $\nu_{RF}$ as a function of $\nu_{AO}$ or for two adequate values of $\nu_{AO}$ respectively. These operations, however, do not always determine $m$ unambiguously [11]. Furthermore, the determination is lengthy and, as discussed below, technical problems arise in order to achieve the required accuracy

The basic idea of our work lies in the fact that $z_0$ is of the order of several meters and that our pulsed-FSF laser is the superposition of a continuous field and a pulsed field. At a repetition rate of 2 kHz the pulse width is 32 ns. A fast photodiode (PD2) simultaneously records the two pulses from the two Mach-Zehnder arms. The pulse rise time is fast enough (<1 ns) to determine the time-delay $\tau_p$ between pulses and therefore $z_p = c\tau_p$ with a precision $\delta z_p$ much better than $z_0$ (typically ~ 10 cm). The beat order $m$ is then very quickly determined. This is the integer part of $z_p/z_0$:

$$m = \left\lfloor \frac{z_p}{z_0} \right\rfloor \quad (7)$$

The fractional part of z {z} is measured on the cw overlap of $I_{int}$ (next section) by precise measurement of $\nu_{RF}$ in the $0\text{-}\nu_{RT}$ band. Given the accuracy of $z_p$, it is sufficient to measure the zero order beat frequency in the interval $0\text{-}\nu_{RT}/2$.

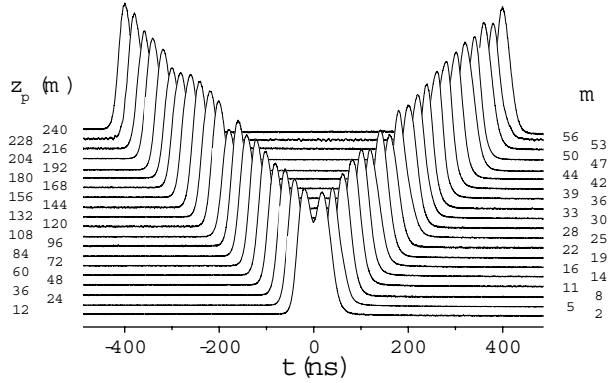

Fig. 6

Figure 6 depicts the dependence of the pulse delay on the number of round trips in the White cell. Pulses are fitted to a sum of two Gaussians, which yield $z_p$ (left scale) with a standard deviation of about 10 cm. $m$ is then determined (right scale). The optical path difference obtained with the White cell varies from 12 m to 240 m in steps of about 12 m. Below 12 m, measurements are limited only by the temporal width of our pulsed pump laser. If necessary, a nanosecond pulsed pump laser should solve this problem easily.

The proposed method seems to fail when $z$ is equal to $mz_0$ within $\delta z_p$. In that case the uncertainty in the measurement of $m$ is one. Then, the unambiguous determination of $m$ requires a second measurement. Since $z$ is the path difference between the two arms of the interferometer, a second measurement can be done by increasing (or decreasing) the length of the reference arm by an amount greater than $2\delta z_p$. Combining these two measurements allows one to determine $m$ unambiguously. A double-reference arm is easy to implement. This double-reference arm method works as long as $\delta z_p$ is much smaller than $z_0$. Otherwise it is possible to use a narrower pulsed pump laser (picosecond rise time have been demonstrated with dye lasers).

Finally note that the AO modulator could also be gated, but the build-up of the FSF laser is too slow [8] to give an unambiguous measurement of *m*.

### *3.4 Long range measurements and coherence*

In the moving-comb model, when the low-pass band is limited to $\nu_{RT}$, the RF-beat corresponds to beating between modes separated by $\Delta\nu = \Delta m \times \nu_{RT}$, where $\Delta m$ is the mode number difference. As we show below, $\Delta\nu$ may be greater than the width of the optical spectrum $\Delta\nu_L$. In this work we studied experimentally the transition where $\Delta\nu$, initially less than $\Delta\nu_L$, becomes greater. Our FSF laser can have two line widths: 4.5 GHz with an intra-cavity etalon [8] and 30 GHz without etalon (measured respectively with a high resolution Fabry-Perot and a SOPRA 2000 monochromator). In our experimental set-up, the distance interval of 12-240 m corresponds to the RF-beat between modes separated by $\Delta m = 2\text{-}56$, that is to say, a frequency difference $\Delta\nu$ ranging from 0.25 GHz to 15 GHz. With a width of the optical spectrum of 4.5 GHz and distances increasing from few meters up to 240 m, the transition can easily be crossed. The path difference corresponding to this transition is:

$$z_t = \frac{\Delta\nu_L}{\nu_{RT}} z_0 \qquad (8)$$

The value of $z_t$ is 508 m and 76 m for the width of the optical spectrum of 30 GHz and 4.5 GHz respectively.

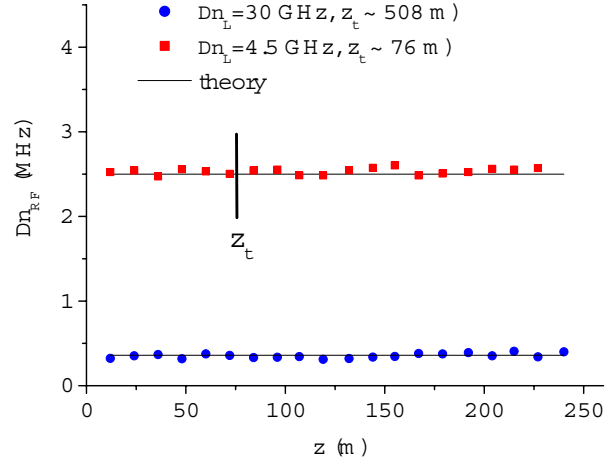

Fig. 7

Figure 7 shows the dependence of the RF-beat width $\Delta\nu_{RF}$ on z, for both width of the optical spectrum. It remains constant without anomaly through the transition. Throughout the path difference range, it is in very good agreement with the calculated value [11]:

$$\Delta\nu_{RF} = \frac{4\sqrt{2}}{3\pi} \frac{1}{\Delta\nu_L} 2\nu_{AO}\nu_{RT} \qquad (9)$$

It is interesting to note that the measurement of $\Delta\nu_{RF}$ also provides a simple and accurate method for determining the width of the optical spectrum of a FSF laser. $2\nu_{AO}$, $\nu_{RT}$ and $\Delta\nu_{RF}$ are measured directly in the FFT of $I_{int}(t)$. Experimentally, we find a line width of 4.2 GHz when the etalon is used. In reference [8] a width of 4.5 GHz was measured using a Fabry-Perot interferometer of high finesse, which is difficult to align.

In the framework of the moving comb model, this experimental result shows that two modes separated by 15 GHz (z ~ 240m) interfere coherently while the width of the optical spectrum is only 4.5 GHz. It has even been shown in the literature [7] that two modes, separated by $\Delta m = 30\,000$, still interfered coherently, without degradation of the spectral width. The mechanism of this long range correlation has never been interpreted within the moving-comb model. The Yatsenko model does not need to invoke the beat between widely separated modes: the presence of the doublets $\pm \nu_{RF}$ is sufficient to interpret our results.

## 4. Precision

It is interesting to compare the OFDR and FSF laser interferometry techniques. OFDR uses a frequency scanned single-mode laser. Today, OFDR is no longer limited by the coherence length of the laser. Commercial lasers with line widths below 1 MHz yield coherence lengths of several hundred meters. The limitation comes from the technically available chirp rate and from the lack of linearity of scanning. Moreover, in OFDR the maximum measurable distance $z_{max}$ and absolute accuracy $\delta z$ are linked by the relation (assuming a perfectly linear scan) [19]:

$$z_{\max} \delta z = \frac{c^2}{\gamma} \quad (10)$$

where $\gamma$ is the chirp rate. For example, a distance of 1000 m measured with an accuracy of $10^{-7}$ needs a chirp-rate of the order of GHz/ns. The frequency chirp is obtained by mechanical, thermal or electronic means. Such chirp rates would require scanning the laser output mirror by a few micrometers within a nanosecond, which has not until now been possible. Moreover, linearity is a real additional problem. Nonetheless, this technique has a proven precision of few tens of microns for short distances (<1m) and a precision of a few meters [20, inner ref] for long distances.

Using an FSF laser changes the issue radically. The chirp-rate can be colossal and the chirp perfectly linear (if the optical cavity length and the acoustic frequency are kept constant). Another remarkable property is the decoupling between precision and maximum distances. It has been demonstrated that the RF-doublets exist for path differences much larger than $c/\Delta\nu_L$, which allows the laser line to be broadened to increase the distance accuracy.

Figure 7 depicts the widths of RF-beats $\Delta\nu_{RF}$ in good agreement with formula (9). A width of 30 GHz can already give a resolution of about 100 microns. As the beat order is determined unambiguously ($dm = 0$), the distance accuracy is given by:

$$\frac{dz}{z} = \frac{d\nu_{AO}}{\nu_{AO}} + \frac{z_0}{z}\frac{\nu_{RF}}{\nu_{RT}}\left(\frac{d\nu_{RF}}{\nu_{RF}} + \frac{d\nu_{RT}}{\nu_{RT}}\right) \qquad (11)$$

As our AO modulator driver is not temperature controlled, the stability of the acoustic frequency $\nu_{AO}$ is only about $3\times10^{-5}$. It is then measured in real time with a high resolution frequency counter having an accuracy of $2\times10^{-7}$. $\nu_{RF}$ and $\nu_{RT}$ are measured in the FFT of $I_{int}$ with the oscilloscope (TDS 3034B), which has an accuracy of $2\times10^{-5}$.

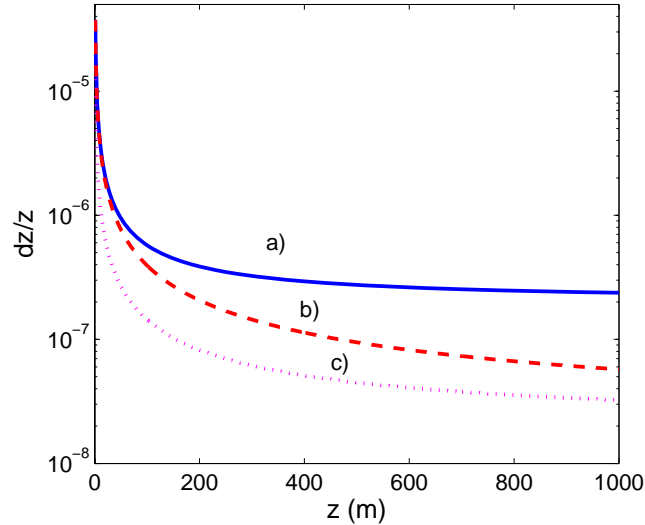

Fig.8

Figure 8 shows the dependence of the precision of our measurements on the distance $z$ (curve a of Fig. 8). An accuracy of $10^{-6}$ is achieved for long distances $z > 50$ m. At short distances, the accuracy is dominated by the measurement of $\nu_{RT}$ and $\nu_{RF}$. At long distances, it is dominated by $\nu_{AO}$. Techniques named Oven Controlled Crystal Oscillator (OCXO) and Rubidium Crystal Oscillator Controlled (RCBC) can easily stabilize the RF frequency applied to the AO with accuracies ranging from $10^{-6}$ to $10^{-10}$. With an acoustic frequency accuracy of $2\times10^{-8}$ (commercial), an accuracy of $10^{-7}$ is reached for distances above half a kilometer (curve b of Fig. 8). At shorter distances, the accuracy of $\nu_{RT}$ and $\nu_{RF}$ must also be improved. It is not enough to increase the precision of the spectrum analyzer (temperature stabilization ...), since the laser line must also be broadened. In our laser configuration, broadening by a factor 3 is possible. The goal of an accuracy of $10^{-7}$ is achievable for distances greater than tens of meters with a laser line-width of about 100 GHz (curve c of Fig. 8). Yatsenko et al. [21] have showed by injecting a single-mode laser that higher signal to noise can be obtained. However, for accuracies better than $10^{-7}$, it becomes necessary to measure independently the index of refraction of the air. Methods using frequency combs from a femtosecond laser are under development [22]. For long distance measurements, the power of FSF laser interferometry lies in the fact that the accuracy is essentially a matter of the electronic limitations of the AO driver, which can be easily solved for a few thousand dollars.

## V Conclusion

The output signal of a Mach-Zehner injected by an FSF laser allows measurement of long distances $z$:

$$z = mz_0 + \{z\} \qquad (12)$$

The beat-order $m$ and the fractional part $\{z\}$ must be determined separately. The interest of our pulsed-FSF laser lies in the superposition of a pulsed and a cw field. The first determines $m$ unambiguously and the second determines the fractional part $\{z\}$ with precision. $z_0$ is a metric that depends only on the acoustic frequency $\nu_{AO}$, which may be very accurate. In this work, $z_0$ is approximately 4.29 m, but, in the configuration of reference [12], which uses two AOs used in substrative-frequency, $z_0$ can be much larger. Typically with $\nu_{AO} = 0.2$ MHz, $z_0$ is about 750 m. Up to this distance, measurement of the zero order RF-beat is sufficient. However, it requires a precision of the frequency $\nu_{AO}$ that is at least two orders of magnitude better than that considered in Figure 8. In that case, to reach an accuracy of $10^{-7}$ the stabilization technique RCBC then becomes necessary.

This work also shows that the moving-comb picture should be abandoned because interferometric long-range measurements lead to a paradox that has no physical explanation: two modes cannot interfere unless they coexist .

## Acknowledgments

We thank the LIPhy laboratory and the CNRS for funding this project. The author is very grateful to E. Geissler for his contribution in reading the paper.

**Figure captions**

1. Fig. 1. Experimental set-up. Pulsed-FSF laser injected in a Mach-Zehnder interferometer with adjustable long-arm (12 m – 240 m). Filtered and amplified PD1 and PD3 signals give the FSF and the interferometric spectra respectively. PD2 signal allows the beat-order determination.

2. Fig. 2. Experimental intensity (a), RF spectra (b) and (c) of the FSF laser. The high speed amplifier cut low frequencies between DC and 50 kHz. $2\nu_{AO}$ beat is a residual acoustic modulation. $\nu_{RT}$ is the free spectral range of the cavity. Inset (c) is in linear scale.

3. Fig. 3. Experimental intensity (a) and RF spectra (b) of the Mach-Zehnder output for $z = 108$ m. The oscilloscope bandwidth is 300 MHz.

4. Fig. 4. (color in print and on the Web) Beat frequency intensities versus distance $z$ and frequency $\nu$ in the moving-comb model without amplitude modulation ($A=0$). $\nu_{RT}$ recurrences are fixed and $\nu_{RF}$ doublets evolve linearly with $z$. Note over intensities at $\nu_{RF} = k \times 2\nu_{AO}$.

5. Fig. 5. (color in print and on the Web) Beat frequency intensities versus distance $z$ and frequency $\nu$ in Yatsenko model.

6. Fig. 6. Output pulses from the two arms of the Mach-Zehnder interferometer for path-differences ranging from ~ 12 m to ~240 m. The beat order $m$ is deduced from the delay time.

7. Fig.7. Evolution of the RF width for $\Delta\nu_L = 4.5$ GHz and $\Delta\nu_L = 30$ GHz. The $z_t$ transition shows no anomaly and no width evolution.

8. Fig. 8. Precision versus distance: a) $v_{AO}$ accuracy $2\times10^{-7}$, $v_{RT}$ and $v_{RF}$ accuracy $3\times10^{-5}$, $\Delta v_L = 30$ GHz, b) $v_{AO}$ accuracy $2\times10^{-8}$, $v_{RT}$ and $v_{RF}$ accuracy $3\times10^{-5}$, $\Delta v_L = 30$ GHz, c) $v_{AO}$ accuracy $2\times10^{-8}$, $v_{RT}$ and $v_{RF}$ accuracy $1\times10^{-5}$, $\Delta v_L = 100$ GHz.